\documentclass[prc,10pt,twocolumn,showpacs,amssymb,aps,nofootinbib,floatfix]{revtex4-1}
\usepackage{epsfig,color,amsmath}

\newcommand{\MeV}{\hbox{MeV}}
\newcommand{\GeV}{\hbox{GeV}}
\newcommand{\fm}{\hbox{fm}} 
\newcommand{\dd}{\mathrm{d}}

\begin{document} 
\hbadness=10000
\topmargin -0.8cm\oddsidemargin = -0.7cm\evensidemargin = -0.7cm

\hyphenation{know-ledge
re-so-nan-ces}

\title{The influence of bulk evolution models on heavy-quark phenomenology}

\author{P.~B.~Gossiaux$^{1}$, S.~Vogel$^{1}$, H.~van~Hees{${}^{2}$},
  J.~Aichelin$^{1}$, R.~Rapp{${}^3$}, M.~He{${}^3$}, M.~Bluhm$^{1}$} 
\email{Corresponding authors: 
pol.gossiaux@subatech.in2p3.fr, svogel@subatech.in2p3.fr}
  \affiliation{${}^{1}$SUBATECH, UMR 6457,
  Laboratoire de Physique Subatomique et des Technologies Associ\'ees \\
  University of Nantes - IN2P3/CNRS - Ecole des Mines de Nantes \\
  4 rue Alfred Kastler, F-44072 Nantes Cedex 03, France; \\
  ${}^2$Institut f{\"u}r Theoretische Physik, Universit{\"a}t Giessen,
  Heinrich-Buff-Ring 16, D-36392 Giessen, Germany; \\
  ${}^3$Cyclotron Institute and Department of Physics and Astronony,
  Texas A{\&}M University, College Station, Texas 77843-3366, USA}
\date{\today}  

\begin{abstract}
  We study the impact of different Quark-Gluon Plasma expansion
  scenarios in heavy-ion collisions on spectra and elliptic flow of
  heavy quarks. For identical heavy-quark transport coefficients
  relativistic Langevin simulations with different expansion scenarios
  can lead to appreciable variations in the calculated suppression and
  elliptic flow of the heavy-quark spectra, by up to a factor of two. A
  cross comparison with two sets of transport coefficients supports
  these findings, illustrating the importance of realistic expansion
  models for quantitative evaluations of heavy-quark observables in
  heavy-ion collisions. It also turns out that differences in freeze-out
  prescriptions and Langevin realizations play a significant role in
  these variations. Light-quark observables are essential in reducing
  the uncertainties associated with the bulk-matter evolution, even
  though uncertainties due to the freeze-out prescription persist.
\end{abstract}
\pacs{14.65.Dw, 25.75.Ld, 24.10.Nz, 24.85.+p}

\maketitle

\section{Introduction}

One of the striking discoveries of the heavy-ion program at the
Relativistic Heavy Ion Collider (RHIC) is that the medium created in
$200 \; A \GeV$ Au-Au collisions behaves like a nearly perfect liquid
\cite{Adams:2005dq,Adcox:2004mh,Back:2004je,Arsene:2004fa,Zajc:2007ey,Muller:2007rs,Vogel:2007yq,Cassing:2008sv}.
To further investigate this new state of matter, usually referred to as
the Quark Gluon Plasma (QGP), penetrating and well calibrated probes are
essential to quantitatively deduce the effect of the medium on those
probes. One of these probes are heavy quarks. The experimental
investigation of heavy-quark probes at RHIC has been ongoing for several
years now~\cite{Abelev:2006db,Adare:2006nq}. Two rather unexpected
observations have emerged.  Heavy mesons, despite their large mass,
exhibit (a) an elliptic flow comparable to that of light mesons,
implying collective motion of heavy quarks in the expanding medium, and,
(b) a suppression at high transverse momentum ($p_T$) similar to light
mesons, implying a substantial energy loss of fast heavy quarks while
traversing the medium.

Several theoretical approaches have been put forward to describe
heavy-quark energy loss, or, more generally, heavy-quark diffusion, in
the
QGP~\cite{Svetitsky:1987gq,vanHees:2004gq,Moore:2004tg,Gossiaux:2004qw,Mustafa:2004dr,vanHees:2005wb,Armesto:2005mz,Gossiaux:2006yu,vanHees:2007me,Molnar:2004ph,Zhang:2003wk,Linnyk:2008hp,
  Gossiaux:2008jv,Gossiaux:2009hr}. Two basic ingredients are required
to perform quantitative calculations for the modifications of the
initial spectra. On the one hand, one needs a good knowledge of the
microscopic interactions of heavy quarks in the plasma, as encoded in
their transport properties (drag and diffusion coefficients). On the
other hand, one needs a realistic description of the expanding medium
through which the heavy quarks propagate. The experimental spectra of
heavy hadrons (and their decay electrons) only reflect a combination of
both components. This may lead to ambiguities in disentangling the two
ingredients.

A closer inspection of the theoretical modeling reveals differences in
both macroscopic expansion scenarios and microscopic transport
coefficients. A first comparison of these issues has been carried out in
Ref.~\cite{Rapp:2009my}. It is the purpose of the present article to separate 
both ingredients and study the influence of the medium description, the
freeze-out prescription and the Langevin realization on the heavy-quark
spectra. Here, we focus on two previously used medium
descriptions, namely the hydrodynamical model by Kolb and
Heinz~\cite{Kolb:2003dz}, and the more schematic elliptical fireball
model by van Hees et al.~\cite{vanHees:2004gq,vanHees:2005wb}.

The objective of this work is not to reproduce experimental data for
heavy-flavor observables. Rather, we employ a common model for the
elementary interaction but change the medium through which the heavy
quarks propagate, or alternatively use a common description of the
medium and change the elementary interaction, to elucidate the influence
of both ingredients on the final spectra.

Our article is organized as follows: In Sec.~\ref{sect:medium} we
describe the two medium descriptions used in our comparison and also
briefly recall the main ingredients to each of the two transport
calculations~\cite{vanHees:2004gq,vanHees:2005wb,Gossiaux:2008jv,Gossiaux:2009hr}.
In Sec.~\ref{sect:HQ-spec} we calculate the spectra for different
combinations of transport coefficients and medium descriptions. In this
way we can separate the influence of the expansion scenario from that of
the elementary interaction of the heavy quarks with the plasma
constituents. For definiteness, this comparison is conducted with the
numerical (Langevin) implementation used in Ref.~\cite{Gossiaux:2008jv},
but we also allude to the Langevin implementation as adopted in
  Ref.~\cite{vanHees:2005wb}. In Sec.~\ref{sect:v2-light} we elaborate
on how light-quark observables, like the elliptic flow of pions, are
constrained within the two expansion scenarios and can help to
distinguish the latter. We conclude in Sec.~\ref{sect:concl}.

\section{Medium descriptions and transport coefficients}
\label{sect:medium}

Let us start by recalling the basic features of the medium expansion in
Ref.~\cite{vanHees:2005wb}, where an elliptical fireball has been used
to model the evolution of the medium created in $b=7 \; \fm$ Au+Au
collisions at $\sqrt{s}=200\,{\rm GeV}/c$, referred to as the vHR (van
Hees/Rapp) medium in the following. It employs a QGP equation of state
with $N_f=2.5$ effective flavors which yields an entropy density
\begin{equation}
s_{\text{QGP}}(T) = \frac{S}{V(t)} = \frac{4\pi^2}{90}(16+10.5N_f)T^3 , 
\end{equation}
with T being the temperature. The total entropy ($S\simeq4600$ in
$\Delta y=1.8$ units of rapidity) is assumed to be time independent. The
fireball volume is parametrized as a function of time according to
\begin{equation}
V(t)=\pi a(t) b(t) (z_0 + c t), 
\end{equation}
with 
\begin{equation}
\begin{split}
  a(t)=a_0 &+ v_{\infty} \left[ t - \frac{1-\exp(-At)}{A} \right] \\ 
  &- \Delta v \left[ t - \frac{1-\exp(-Bt)}{B} \right],  \\
  b(t)=b_0 &+ v_{\infty} \left[ t - \frac{1-\exp(-At)}{A} \right] \\ 
  &+ \Delta v \left[ t - \frac{1-\exp(-Bt)}{B} \right] \ .
\end{split}
\label{ab}
\end{equation}
The tuning of the parameters,
\begin{equation}
\begin{array}{lll}
  a_0=5.562 \; \fm,
  &
  A=0.55\,c/\fm,
  &
  b_0=4.450\; \fm,\\[3mm]
  B=1.3\,c/\fm,
  &
  v_\infty=0.52 c, & \Delta v=0.122 c \ , \\
\end{array}
\end{equation}
will be discussed below. Since the fireball is approximated as
homogeneous, the thermodynamic variables only depend on time but not on
position. Therefore, for a given volume evolution and total entropy, the
temperature, pressure and energy density can be calculated as a function
of time as
\begin{alignat}{2}
p &=\frac{\pi^2}{90}(16+10.5N_f)T^4-B_{\text{QGP}},\\
\epsilon &= Ts-p = 3 \cdot \frac{\pi^2}{90}(16+10.5N_f)T^4+B_{\text{QGP}},
\end{alignat}
where the bag constant, $B_{\text{QGP}}=356\; \MeV/\fm^3$, ensures the
continuity of pressure through the mixed phase (as in the EOS-Q of
Refs.~\cite{Kolb:2000sd,Kolb:2003dz}). The phase transition is modeled
by a standard mixed-phase construction at constant temperature, $T_c=180
\;\MeV$, with critical QGP and HG energy densities of
$\epsilon_c^{\text{QGP}}=2.25 \; \GeV/\fm^3$ and
$\epsilon_c^{\text{HG}}=0.82 \; \GeV/\fm^3$ (the latter following from a
thermal hadron-resonance-gas model). The expansion parameters in
Eq.~(\ref{ab}) have been determined to mimic the time evolution of the
hydrodynamical model of Ref.~\cite{Kolb:2000sd} but with final
quark-momentum spectra and elliptic flow ($v_2$) adjusted to results of
coalescence-model fits to empirical pion and kaon
spectra~\cite{Greco:2003mm}. Note that this absorbs the effects of a
subsequent hadronic evolution. The effective quark mass has been set to
$m_q=0.3 \; \GeV$, and the freeze-out prescription has been chosen
consistent with the equilibrium limit (i.e., long-time limit) of the
post-point Ito realization of the Langevin simulation for the heavy
quarks within the vHR model,
\begin{equation}
f_{\rm ML}(x,\vec{p})= 
\frac{1}{(2 \pi)^3} \frac{p \cdot u(x)}{E} \exp \left
 [-\frac{p \cdot u(x)}{T(t)} \right ]
\label{eq_milekhin_dist}
\end{equation}
(cf. Appendix~\ref{app:langevin-freeze-out} for details). This
function is reminiscent of the one obtained within the so-called 
Milekhin freeze-out prescription~\cite{Russkikh:2006aa,Ivanov:2008zi}
and will therefore be referred to as ``Milekhin-like'' in the following.
In Eq.~(\ref{eq_milekhin_dist}), $u(x)$ is the four-velocity vector
field of the medium. In vHR the transverse flow field,
$\vec{v}_\perp(t,\vec{x})$, is constructed with help of confocal
elliptical coordinates in the transverse $(x,y)$ plane, i.e.,
\begin{equation}
\vec{v} = \left(\frac{r}{r_B} v_b(t) \cos v, \frac{r}{r_B} v_a(t) \sin v,0\right),
\label{eq_velocity}
\end{equation}
where $v_{a}(t)=\dot{a}(t)$, $v_{b}(t)=\dot{b}(t)$, and
\begin{eqnarray}
\vec{r}&\equiv&(x,y)=(0.6 a_0 \sinh u \cos v, 0.6 a_0 \cosh u \sin v)
\nonumber\\
\vec{r}_B&\equiv& (b(t) \cos v, a(t) \sin v)\,.
\end{eqnarray}
The expansion parameters in Eq.~(\ref{ab}) produce final light-quark
spectra with an average surface-flow velocity of $v_{b}(t_{\text{mix}})
\simeq 0.55$ and total elliptic flow of $v_2 \simeq 5.6\%$, at the end
of the mixed phase. Note that these parameters are designed to reproduce
hadron observables, not necessarily the result of a hydrodynamic model
at the end of the mixed phase.

The second model to describe the expansion of the medium is (2+1)D
hydrodynamical calculation by Kolb and Heinz (labeled
KH)~\cite{Kolb:2003dz}. The equations underlying ideal hydrodynamics are
the conservation of energy and momentum and appropriate currents (e.g.,
baryon number),
\begin{equation}
\label{eq1}
\partial_\mu T^{\mu \nu}(x)=0\,, \quad 
\partial_\mu j^{\mu}(x)=0.
\end{equation}
The equation of state is very similar to vHR, consisting of two
parts. The hadronic phase is described as a gas of non-interacting
hadronic resonances,
containing essentially the same resonances as in the vHR fireball model
but using a lower critical temperature, $T_c=164 \; \MeV$. Above the
phase transition, in the QGP phase, the system is modeled as a
non-interacting gas of $u$, $d$, $s$ quarks and gluons, with an external
bag pressure (of a similar value as the one used in vHR). The temperature 
and energy density are local quantities in this approach. For more information 
on the modeling of the system with this approach the reader is referred to
Ref.~\cite{Kolb:2003dz}.

The parameters of both medium descriptions have been fixed to describe
Au+Au collisions at $\sqrt{s}=200 \; A \GeV$ at an impact parameter of
$b=7~\mathrm{fm}$. Note that the hydrodynamical model describes the
hadron-$p_T$ spectra and $v_2$ at thermal freeze-out in the hadronic
phase, while the vHR fireball has been tuned to reproduce quark-momentum
distributions at the end of the mixed phase such as to reproduce
hadronic spectra within a coalescence model for
hadronization~\cite{Greco:2003mm}. Thus, by construction, the elliptic
flow, $v_2$, in the vHR scenario is larger than in the KH scenario.

\begin{figure}[htb]
\epsfig{width=8cm,clip=1,figure= 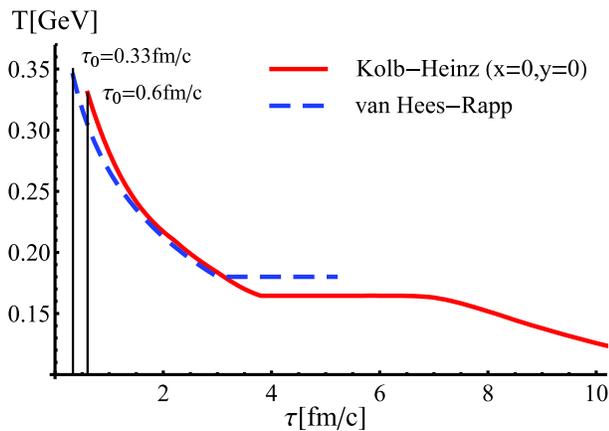}
\caption{(color online) Temperature profiles for KH
  hydrodynamics (central cell) and for the vHR  elliptical
  fireball. One observes different lifetimes of the QGP and 
  mixed phases for both approaches. Additionally, the hadronic phase of
  the KH description is depicted but heavy meson interactions are neglected. }
\label{temp_profiles}
\end{figure}

In Fig.~\ref{temp_profiles} we display the temperature-time profiles
extracted from KH hydrodynamics and the vHR fireball. In the
hydrodynamical simulation the temperature is local, and for this
comparison we take the temperature in the central cell. The temperature
in the elliptic fireball depends only on its volume and is equal at all
points in space. The vHR medium prescription starts at $\tau_0 = 0.33 \;
\fm/c$ and stops at roughly $5 \; \fm/c$ with the end of the mixed
phase. For the KH prescription the medium evolution lasts longer,
starting at $\tau_0 = 0.6 \; \fm/c$ and ending the mixed phase at $\sim
7 \;\fm/c$, followed by a hadronic phase. The impact of the hadronic
phase on the heavy-quark energy loss has been neglected when using the
KH medium.

\begin{figure}[htb]
\epsfig{width=8cm,clip=1,figure= 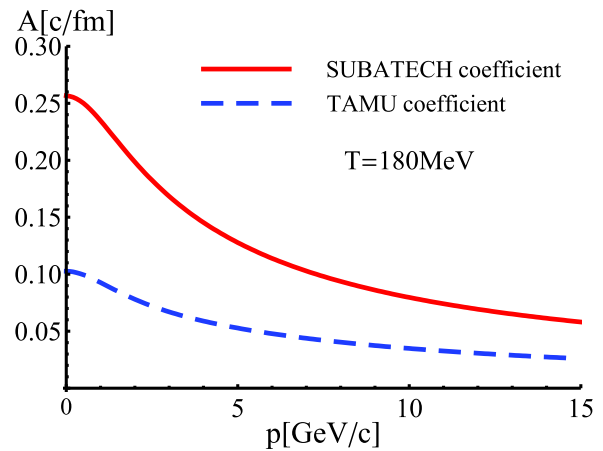}
\caption{(color online) Drag coefficients for pQCD with running $\alpha_s$
(solid line) and for resonance+pQCD interactions (dashed line), as a 
function of momentum for a temperature of 180~MeV. A clear difference 
between the two approaches is observed.}
\label{drag_diff_coeff}
\end{figure}

Next we turn to the different descriptions of heavy-quark diffusion. In
Refs.~\cite{Gossiaux:2008jv,Gossiaux:2009hr} a pQCD calculation with
running coupling has been used, whereas in Ref.~\cite{vanHees:2005wb}
matrix elements based on an effective resonance
model~\cite{vanHees:2004gq} have been employed. For this study, the
interactions are implemented through the use of a Langevin-transport
model (cf. Appendix~\ref{app:langevin-freeze-out} for more details)
characterized by ensembles of random momentum kicks,
\begin{equation}
\dd p_j=-\Gamma(t,\vec{p}) p_j \dd t + \sqrt{\dd t} C_{jk}(t,\vec{p}+\xi \dd
\vec{p}) w_k \ ,
\end{equation}
which shift the heavy-quark momentum at each time step, $\dd t$; the
first term, $-\Gamma(t,\vec{p}) p_j$, is a friction term, $w_k$ are
Gaussian-distributed random variables and the $C_{jk}$ are related to
the diffusion tensor, $\hat{B}$ (see appendix). The choice of $\xi \in
\{0,1/2,1 \}$ characterizes realizations of the Langevin process, known
as pre-point Ito, Stratonovich and post-point Ito (or
H{\"a}nggi-Klimontovich) relations, respectively. For a particle of
mass, $M$, which is large compared to the temperature of the ambient
medium, one has $\Gamma=A+O(T/M)\approx A$ independent of the
realization, where $A$ is the drag coefficient responsible for the
energy loss of particle, defined from the microscopic interaction by
\begin{equation}
\begin{split}
\label{drag-coefficient}
A &= \frac{1}{2pE} \int \frac{d^3k}{(2\pi)^3 2k}
\int\frac{d^3k'}{(2\pi)^3 2k'} \int \frac{d^3p'}{(2\pi)^3 2E'} n_i(k) \\
& \times (2\pi)^4\delta^{(4)}(p\!+\!k\!-\!p'\!-\!k')\frac{1}{d_i}
\sum \left|{\cal M}_i\right|^2\, (p-p') \ .
\end{split}
\end{equation}
Figure~\ref{drag_diff_coeff} depicts the drag coefficients computed in
terms of the scattering-matrix elements of the underlying microscopic
model for the heavy-quark interaction as a function of three-momentum at
a fixed temperature of $T=180 \; \MeV$ for the resonance model of
Ref.~\cite{vanHees:2004gq} and for the pQCD model of
Refs.~\cite{Gossiaux:2008jv,Gossiaux:2009hr} (dubbed TAMU and SUBATECH
coefficients, respectively). One finds sizable differences by more than
a factor of two.

\section{Systematic comparison of heavy-quark spectra}
\label{sect:HQ-spec}

In this section we analyze heavy-quark spectra and elliptic flow by
combining both medium descriptions with both transport coefficients, as
alluded to in the previous section. In particular, we also study the
time evolution of the elliptic flow. This requires the identification of
a suitable variable characterizing the time evolution. In the vHR medium
the energy density and the evolution time are uniquely correlated, but
such a definitive relation does not exist in the hydrodynamic
approach. In the KH medium we terminate the interactions of a heavy
  quark in a fluid cell as soon as its energy density falls below the
freeze-out energy density, $\epsilon_{\rm fo}$, and evaluate the
heavy-quark spectra on the pertinent hyper-surface.

\begin{figure}[htb]
\epsfig{width=8cm,clip=1,figure= 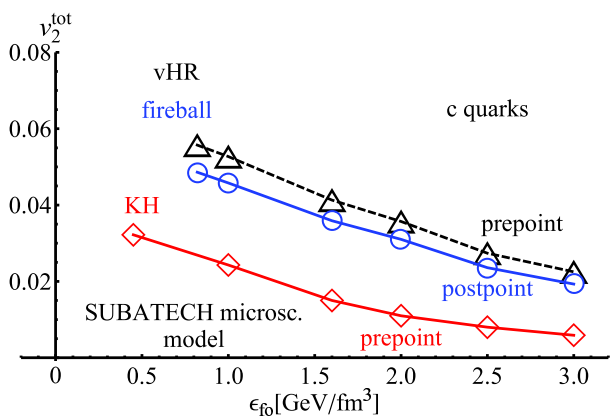}
\caption{(color online) Total $v_2$ of  charm quarks as a function of 
  freeze-out energy density for two different medium descriptions with identical
  (pre-point) Langevin realization (triangles and diamonds) and for the vHR
  evolution with post-point Langevin (circles).}
\label{inclusive_v2}
\end{figure}

In Fig.~\ref{inclusive_v2} we show the total $v_2$ for
$c$-quarks, defined by
\begin{equation}
  v_2^{\rm tot}:=\frac{\int p_T \mathrm{d}p_T \mathrm{d}\varphi \cos(2\varphi) 
    \frac{\dd^3N}{\mathrm{d}y p_T \mathrm{d}p_T \mathrm{d}\varphi}}
  {\int p_T \mathrm{d}p_T \mathrm{d}\varphi
    \frac{\mathrm{d}^3N}{\mathrm{d}y p_T \mathrm{d}p_T
      \mathrm{d}\varphi}} \ , 
\label{eq_def_incl_v2}
\end{equation} 
as a function of different freeze-out energy densities, $\epsilon_{\rm
  fo}$, down to the last point, where still a fraction of QGP exists,
$\epsilon_{\rm fo}=0.45\,{\rm GeV}/{\rm fm^3}$ in KH and $\epsilon_{\rm
  fo}=0.82\,{\rm GeV}/{\rm fm^3}$ in vHR. All curves are calculated with
the same microscopic model for the heavy-quark transport coefficients
(pQCD with running
coupling)~\cite{Gossiaux:2008jv,Gossiaux:2009hr}\footnote{with the
  longitudinal coefficient $B_L$ imposed as $B_L=T E A$ in order to
  enforce the Einstein relation.}, but with different medium evolutions
and Langevin realizations. The full red line (with diamonds) depicts the
KH hydro evolution with the pre-point Langevin, the dashed line (with
triangles) the equivalent result for the vHR fireball, and the full blue
line (with circles) the vHR medium with a post-point Langevin; the
latter corresponds to the results of Refs.~\cite{vanHees:2005wb}.

We first note that the precise realization of the Langevin process has a
moderate influence on the outcome of the calculation in the heavy-quark
sector. For the two different medium descriptions, the discrepancies are
larger, up to a factor of 2 at a given energy density. This suggests
that both media carry different momentum eccentricities along their
evolution\footnote{This feature, which we later on refer to as
  ``intrinsic $v_2$'' of the medium, could be further quantified by,
  e.g., evaluating the anisotropy, $\epsilon_p$, of the energy-momentum
  tensor.}, which, in turn, are transferred to the heavy-quark motion.
We already pointed out that the vHR evolution has been tuned to the
empirical $v_2$ of light quarks at the end of the mixed phase. We will
return to this question in Sec.~\ref{sect:v2-light} below; apparently,
the moderate difference between $5.6\%$ (vHR) and $4.8\%$ (KH hydro) is
not the main cause for the effect seen in Fig.~\ref{inclusive_v2}.
  
In the following, we elaborate on how the differences in the medium
expansion affect heavy-quark spectra by performing a comparison with
different transport coefficients and medium prescriptions. Specifically,
we compare two basic features of heavy-quark spectra at RHIC, i.e., the
nuclear suppression factor, $R_{AA}(p_T)$, and the elliptic flow,
$v_2(p_T)$, by studying four setups:
\begin{enumerate}
\item[I.] KH medium / SUBATECH coefficients 
\item[II.] KH medium / TAMU coefficients
\item[III.] vHR medium / SUBATECH coefficients
\item[IV.] vHR medium / TAMU coefficients
\end{enumerate}
The scenario described as ``SUBATECH coefficients'' are those published
in Refs.~\cite{Gossiaux:2008jv,Gossiaux:2009hr}, whereas the drag and
diffusion coefficients used in
Refs.~\cite{vanHees:2004gq,vanHees:2005wb} are labeled as ``TAMU
coefficients''. The scenarios I and IV describe the published data on
the nuclear suppression factor, $R_{AA}$, of semileptonic electrons
fairly well, despite of different assumptions on the medium and the
transport coefficients. To disentangle the effects of the two
components, we swap the medium description with the description of the
heavy-quark interaction (scenarios II and III). In this way we can
cross-compare the results and better identify changes due to the medium
or the diffusion mechanism. Since the precise realization of the
Langevin process has rather little impact in the heavy-quark sector, we
have chosen to proceed with the pre-point (post-point) prescription for
the KH (vHR) medium; the motivation for this choice will be justified in
Sec.~\ref{sect:v2-light}.

\begin{figure}[!t]
\epsfig{width=8cm,clip=1,figure= 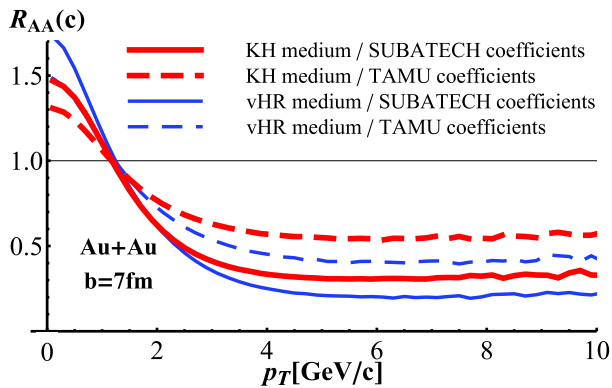}
\caption{(color online) Nuclear modification factor, $R_{AA}$, of charm
  quarks at RHIC energies for the four different scenarios. The lines
  are identified as scenario I (full/thick red), scenario II
  (dashed/thick red), scenario III (full/thin blue) and scenario IV
  (dashed/thin blue).}
\label{fig_RAA}
\end{figure}

Figure~\ref{fig_RAA} depicts the nuclear suppression factor of charm
quarks at the end of the respective mixed phases as a function of their
transverse momentum for the four scenarios mentioned above. Scenarios I
and IV give rather similar results, within ca.~20\%. When changing the
medium description and leaving the diffusion mechanism the same, i.e.,
comparing scenarios I with III, or II with IV, one observes a difference
of 30-50\% at high $p_T$ in opposite directions, indicating that the vHR
medium induces more ``stopping'' than the KH medium. As expected, the
smaller friction coefficients of
Refs.~\cite{vanHees:2004gq,vanHees:2005wb} cause a smaller energy loss
than the pQCD+running-$\alpha_s$ model. The maximal deviation of roughly
a factor of 2 occurs when combining the large coefficients with the
``more stopping'' medium (scenario III) compared to the small
coefficients plus ``less stopping'' medium (scenario II).

\begin{figure}[htb]
\epsfig{width=8cm,clip=1,figure= 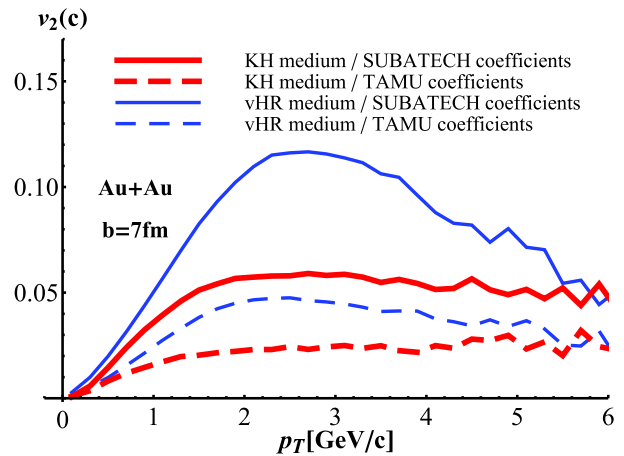}
\caption{(color online) Elliptic flow of charm quarks at RHIC energies.
  The red (blue) lines are computed with the KH hydro (vHR fireball) 
  evolution, and the solid (dashed) correspond to using pQCD+running-$\alpha_s$ 
  (resonance) model transport coefficients. } 
\label{v2}
\end{figure}

Similar features are found when comparing the elliptic flow of charm
quarks, $v_2(p_T)$, see Fig.~\ref{v2}. One observes again that scenarios
I and IV compare reasonably well to each other, on the level of
ca.~20-30\%. However, when interchanging the medium description the
elliptic flow either increases by more than a factor of 2 (scenario III
compared to I) or decreases by roughly a factor of 2 (scenario II
compared to IV) at $p_T \simeq 2$-$4\; \GeV$. The differences at
intermediate $p_T$ are thus stronger for $v_2$ than for $R_{AA}$.

\section{Elliptic flow of light charged particles}
\label{sect:v2-light}

As mentioned above, the calculated $R_{AA}$ and $v_2$ values of charm
quarks cause ambiguities in quantitatively disentangling the effects of
the expansion scenario from that of the microscopic interaction of heavy
quarks in the QGP. It is therefore important to return to light-quark
observables to better distinguish between different
scenarios. Light-hadron observables depend on the final state of the
expansion scenario but are independent of the interaction of heavy
quarks with the plasma constituents. Thus they allow to scrutinize the
description of the medium evolution if the freeze out description were
unique, which is unfortunately not the case.  A key observable is the
elliptic flow of pions or charged particles. In the following we study
in more detail how our two bulk evolution models have been adjusted to
experimental data.

We adopt the standard definition for the differential elliptic flow of a
particle with mass, $m$,
\begin{equation}
v_2(p_T):=\frac{\int \mathrm{d}\varphi \cos(2\varphi) 
 \frac{\mathrm{d}^2N}{p_T \mathrm{d}p_T \mathrm{d}\varphi}}
  {\int \mathrm{d}\varphi 
    \frac{\mathrm{d}^2N}{p_T \mathrm{d}p_T \mathrm{d}\varphi}}\, 
\end{equation}
in terms of the single-particle momentum distribution
function\footnote{The integration over rapidity, $y$, is performed to
  collect the flow of all light partons that are able to interact with
  the heavy quarks in the rapidity interval $\Delta y=1.8$ inherent to
  the fireball.}
\begin{equation}
\frac{\mathrm{d}^2N}{p_T \dd p_T \mathrm{d}\varphi}=\int \mathrm{d}y 
\frac{E \mathrm{d}^3N}{\mathrm{d}^3p}.
\label{eq_spectra_2D}
\end{equation}
If the expanding medium is in local thermal equilibrium during the
expansion the key question is, how to convert the fluid cells,
characterized by a temperature and flow field, into a particle
distribution.

In the KH hydro calculations (as in most other hydrodynamical models)
the Cooper-Frye (CF) prescription~\cite{Cooper:1974mv} is employed to
evaluate the momentum distributions of particles after freeze-out. It
converts a thermal medium instantaneously into a momentum distribution
given by
\begin{equation}
\label{cooper-frye}
\frac{E \; \dd^3N}{\mathrm{d}^3p}= \int \dd \sigma_\mu \; p^\mu f(\vec p,T,u).
\end{equation}
with the Boltzmann - J\"uttner distribution,
\begin{equation}
\label{BJ}
f(\vec p,T,u)=\frac{1}{(2 \pi)^3} \exp \left
    (-\frac{p \cdot u}{T} \right),
\end{equation}
where $T$ and $u$ are the temperature and four-velocity at thermal
freeze-out below which no further interaction occurs, and $\sigma_\mu$
is the hypersurface at constant $T$ (or energy density). In the
following, we will consider this prescription to analyze quark spectra
during the evolution of the QGP as well.

In the vHR fireball model, the Milekhin-like freeze-out prescription has
been adopted. Since the medium is approximated as isotropic, the
hypersurface corresponds to the entire ellipsoid volume defined by the
condition
\begin{equation}
\frac{x^2}{b^2(t)}+ \frac{y^2}{a^2(t)}\le 1 \ ; 
\label{eq_criteria_transv_space}
\end{equation}
The four-velocity at time, $t$, and position, $\vec{x}_\perp$, is
evaluated from Eq.~(\ref{eq_velocity}). According to
Refs.~\cite{Cooper:1974qi,Cooper:1974mv} one thus has
\begin{equation}
\frac{\mathrm{d}^3N}{\mathrm{d}y p_T \mathrm{d}p_T \mathrm{d}\varphi}=
\frac{E\mathrm{d}^3N}{\mathrm{d}^3p}=\int_{\cal V} \frac{\mathrm{d}V}{(2\pi)^3} 
\ {E} \ f_{\mathrm{ML}}\left(p,T,u\right)\,,
\label{dNd3p_CF_fo_labtime}
\end{equation}
with $f_{\mathrm{ML}}$ defined in Eq.~(\ref{eq_milekhin_dist}).  In the
following, we will also evaluate the fireball-$v_2$ with the CF
prescription, by taking $f_{\rm ML}\rightarrow f$ in
Eq.~(\ref{dNd3p_CF_fo_labtime}).

\begin{figure}[htb]
\epsfig{width=8cm,clip=1,figure=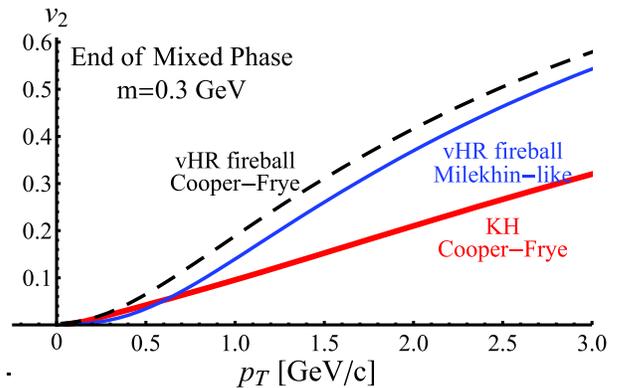}
\caption{(color online) Elliptic flow of particles with mass, $m=300 \;
  \MeV$, as a function of transverse momentum at the end of the
  respective mixed phase for the KH medium with Cooper-Frye freeze-out
  (solid red line) and the vHR medium+Milekhin-like freeze-out (solid
  blue line). The dashed line is the result of applying Cooper-Frye
  freeze-out to the vHR medium.}
\label{fig_v2}
\end{figure}

In Fig.~\ref{fig_v2} we display the differential $v_2$ for constituent
quarks with a mass $m = 300 \; \MeV$, emanating from the KH and vHR
medium, using both the Cooper-Frye and the Milekhin-like descriptions
for the latter. Since the heavy-quark evolutions are terminated in both
media at the end of the mixed phase, we choose the corresponding
hypersurfaces for comparison. It turns out that the KH+Cooper-Frye and
vHr+Milekhin-like prescriptions are rather close up to $p_T \simeq
700\;\MeV$. This encompasses most of the bulk-particles in the medium
(and most of the interactions of the heavy quarks occur with those soft
partons). However, if one applies the CF freeze-out to the vHR fireball
one finds that the parton $v_2$ is systematically above the KH+CF and
the vHR+ML medium, even at low $p_T$ (e.g., by almost a factor of 2 at
$p_T \simeq 700 \;\MeV$, where KH+CF and vHR+ML cross). This reiterates
the evidence found in the context of the charm-quark $v_2$ in
Sec.~\ref{sect:HQ-spec} that the $v_2$-content of the vHR medium is
significantly larger than that of the KH medium.

\begin{figure}[!t]
\epsfig{width=8cm,clip=1,figure= 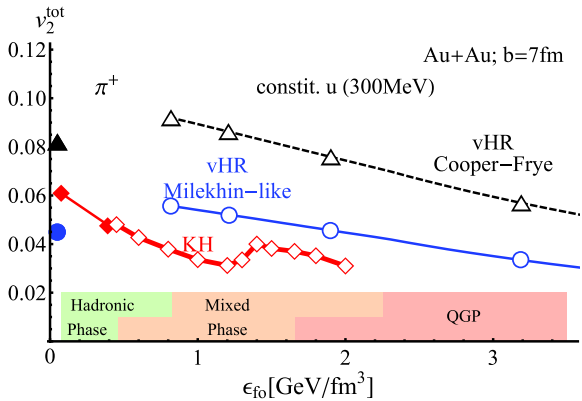}
\caption{(color online) Total $v_2$ as a function of freeze-out
    energy density for constituent quarks ($m=300 \; \MeV$, open
    symbols) and pions ($m=140 \; \MeV$, full symbols). The light quarks
    are evaluated with CF freeze-out for the KH (diamonds) and vHR
    (triangles) medium, as well as with the original Milekhin freeze-out
    for the vHR medium (circles).  The colored boxes at the bottom
    sketch the three stages of the medium evolution (QGP, mixed and
    hadronic phase). The pions have been evaluated in the hadronic phase
    (using Boltzmann statistics, no resonance feeddown) either at the
    end of the mixed phase (vHR medium) or throughout the hadronic
    evolution for the KH medium.}
\label{fig_incl_v2}
\end{figure}

It is furthermore instructive to examine the \emph{total} particle $v_2$
according to Eq.~(\ref{eq_def_incl_v2}) -- averaged over rapidity as in
Eq.~(\ref{eq_spectra_2D}) --, which is particularly suitable to
illuminate time (or energy-density) dependencies. In
Fig.~\ref{fig_incl_v2} we display this quantity as a function of the
freeze-out energy density for $m=300 \; \MeV$ ``partons'' above the
critical value ($0.45 \; \fm^{-3}$ and $0.82 \; \fm^{-3}$ for KH and
vHR, respectively) and for $m=140 \; \MeV$ ``pions'' below.

In the KH medium, for $\epsilon\lesssim 1.4~{\rm GeV}/{\rm fm}^3$, the
direct evaluation of $v_2$ from the physical fields (e.g.
$T_{\mathrm{KH}}(\tau_B,\vec{x}_\perp)$) is hindered by numerical
fluctuations inherent to $\nabla \tau_{\rm fo}(\vec{x}_\perp)$, where
$\tau_{\rm fo}(\vec{x}_\perp)$ is the freeze-out Bjorken time for a
given position in transverse space. We have therefore resorted to a
direct Monte-Carlo sampling of the freeze-out hypersurface, $\Sigma$, in
order to evaluate this quantity in a more robust way.

For the vHR fireball, there is no problem in performing the calculation
for $v_2(\epsilon)$ down to the end of the mixed phase ($\epsilon_{\rm
  fo}=0.82 \; \GeV/\fm^3$). By construction in the original
work~\cite{vanHees:2005wb} the Milekhin-like freeze-out prescription
results in an total $v_2$ of close to 6\%, adjusted to experiment in
semicentral Au-Au collisions at RHIC. The same is true for the
KH-hydro+Cooper-Frye freeze-out, where, however, only 3/4 of the value
is reached at the end of the mixed phase while the remaining 1/4 of the
experimental value develops in the hadronic phase. Applying the CF
freeze-out to the fireball leads to significantly larger values of
ca.~$9\%$ at the end of the mixed phase, which is $\approx 60\%$
($90\%$) larger than the vHR+ML (KH+CF) prescriptions.

Pursuing the bulk evolution beyond the end of the mixed phase, one can
compare the elliptic flow of directly produced pions,
$v_2^{\mathrm{tot}}(\pi^+)$. In the case of the KH hydro, we have
evaluated the $v_2$ directly from the $\pi^+$ spectra resulting from the
hydrodynamical evolution. For the vHR medium, we have proceeded as
described above for partons but simply taken $m=140 \; \MeV$ and a
Boltzmann distribution to resemble pions after the mixed phase. 
The resulting pion $v_2$ at the end of the mixed phase turns out to be 4.5\% (8\%) for
Milekhin-like (CF) freeze-out, which is smaller (larger) than for the KH
medium by roughly 25\% (30\%).

The analysis of the bulk $v_2$ summarized in Figs.~\ref{fig_v2} and
\ref{fig_incl_v2} suggests an explanation for the origin of the
discrepancies found in the charm-quark spectra analyzed in
Sect.~\ref{sect:HQ-spec}: On the one hand, different freeze-out (and
hadronization) prescriptions have been applied in the light-quark sector
for the two medium descriptions, both of which lead to good agreement
with the empirical pion $v_2$. On the other hand, when the two medium
evolutions are analyzed with the same freeze-out prescription, an
appreciable discrepancy in the bulk $v_2$ emerges.  This ``intrinsic''
bulk $v_2$ appears to be the key quantity in communicating the momentum
anisotropy to the heavy quarks propagating through the medium. Maybe
somewhat surprisingly, the results in the heavy-quark sector, for
``realistic'' transport coefficients, are not very sensitive to the
Langevin implementation (i.e., pre-point vs. post-point prescription),
which, in turn, dictated the chosen freeze-out prescriptions in both
media in the first place. Clearly, the theoretical issue remains to
better understand the discrepancies in the different freeze-out
descriptions in connection with the underlying Langevin implementation
of heavy-quark diffusion, which of course should be consistent.

\section{Conclusions}
\label{sect:concl}

We have discussed the impact of the bulk-medium evolution on heavy-quark
phenomenology in heavy-ion collisions by studying two expansion
scenarios, which have been applied earlier to calculate the elliptic
flow and nuclear suppression factor of heavy-quark observables at
RHIC. In both approaches the final results depend on the microscopic
interaction of the heavy quarks with the QGP (as encoded in their
transport coefficients) and on the expansion scenario which provides
different intrinsic $v_2$ values of the bulk medium.  We separated these
ingredients in order to better understand the influence of the medium
descriptions on final spectra. By switching the medium and freeze-out
descriptions one observes differences of around 50\% in the $R_{AA}$ and
$v_2$ of charm-quark spectra. This effect has been cross-checked with
different models for the microscopic input for drag and diffusion
coefficients.

In the past, many efforts have concentrated on a better understanding of
the microscopic interactions between heavy quarks and plasma
constituents. Our study suggests that the influence of different
expansion scenarios on the heavy-quark observables is comparable to that
of different descriptions of the heavy-quark transport coefficients. In
principle, light-quark observables can help to determine the expansion
scenario but differences in freeze-out prescriptions and Langevin
implementations induce significant uncertainties at present. Different
freeze-out descriptions can indeed result in similar values for the
light-meson $v_2$ in models where the ``intrinsic'' elliptic flow of the
hot medium differs appreciably.

\section*{Acknowledgments}

The computational resources have been provided in part by Subatech. We
acknowledge the support by U. Heinz for providing the hydrodynamical
calculations and thank E.~Bratkovskaya and R.J.~Fries for fruitful
discussions and comments. The work of RR and MH has been supported by
the U.S. National Science Foundation under grant no. PHY-0969394 (RR,
MH) and CAREER grant no. PHY-0847538 (MH), and by the A.-v.-Humboldt
Foundation (RR). PBG, SV and JA have been supported by the ANR research
program ``hadrons@LHC'' under grant no.  ANR-08-BLAN-0093-02 and by the
PCRD7/I3-HP program TORIC.

\appendix

\section{Langevin simulations and freeze-out scenarios}
\label{app:langevin-freeze-out}

One of the difficulties in the use of relativistic Langevin simulations
for heavy-quark diffusion processes is the dependence of the resulting
phase-space distribution function on the realization of the stochastic
integral. The Langevin process is defined by the time step,
\begin{equation}
\begin{split}
\label{langevin.1}
\dd x_j &=\frac{p_j}{E} \dd t, \\ 
\dd p_j &=-\Gamma(t,\vec{p}) p_j \dd t + \sqrt{\dd t} C_{jk}(t,\vec{p}+\xi \dd
\vec{p}) w_k \ ,
\end{split}
\end{equation}
for the heavy-quark position and momentum coordinates with respect to
the rest frame of the heat bath. The $w_k(t)$ denote stochastically
independent normally distributed random variables (``white noise''),
\begin{equation}
\label{langevin.2}
\left \langle w_j(t) w_k (t')\right \rangle=\delta(t-t') \delta_{jk} \ .
\end{equation}
As elaborated in Ref.~\cite{Rapp:2009my}, the Langevin process defined
by Eq.~(\ref{langevin.1}) is equivalent to the Fokker-Planck equation
\begin{widetext}
\begin{equation}
\label{langevin.3}
\frac{\partial f}{\partial t} + \frac{p_j}{E} \frac{\partial f}{\partial
  x_j} = \frac{\partial}{\partial p_j} \left [ \left (p_j \Gamma - \xi
    C_{lk} \frac{\partial C_{jk}}{\partial p_l} \right ) f \right ] +
\frac{1}{2} \frac{\partial^2}{\partial p_j \partial p_k} (C_{jl} C_{kl} f) \ ,
\end{equation}
\end{widetext}
for the heavy-quark phase-space distribution function $f$, where one can
identify the usual drag force $A_j=p_j \Gamma - \xi C_{lk}
\frac{\partial C_{jk}}{\partial p_l}$ as well as the diffusion tensor
$\hat{B}=\hat{C}\cdot\hat{C}^T/2$. Here, $\xi \in [0,1]$ defines the
realization of the stochastic integral, incorporating the effects of the
force fluctuations around the average drag or friction force, governed
by the drag coefficient, $\Gamma$. In Ref.~\cite{Gossiaux:2004qw}, a
pre-point Ito prescription has been adopted ($\xi=0$), in which case one
has identically $\Gamma=A$. Furthermore, the diffusion coefficients
$B_L$ and $B_T$ have been adjusted to enforce the Einstein relation by
solving Eq.~(18) of Ref.~\cite{rafelsky:2000} and imposing the
additional constraint that
\begin{equation}
\frac{B_T(p)}{B_L(p)}
=\left(\frac{B_T^{\rm brute}(p)}{B_L^{\rm brute}(p)}\right)^\frac{1}{4} \ ,
\end{equation}
where $B_L^{\rm brute}$ and $B_T^{\rm brute}$ are the coefficients
evaluated directly with help of the differential cross-section for the
microscopic $q/g+Q\rightarrow q'/g'+Q'$ processes. This prescription has
been adopted in order to preserve the anisotropy observed in high-energy
collisions. Once these assumptions were adopted for the case of a fluid
at rest, it has been checked numerically that the equilibrium limit of
the distribution in a moving fluid is compatible with
\begin{equation}
f_{\rm BJ}(\vec{p}_{\rm lab})\propto \exp \left (-\frac{p_{\rm lab}\cdot
    u}{T} \right) \ ,
\end{equation}
where $u=u(t,\vec{x})$ is the four-velocity flow field of the background
medium with respect to the laboratory frame.

In Ref.~\cite{vanHees:2005wb,vanHees:2007me}, following
Ref.~\cite{Moore:2004tg}, the Langevin realization has been chosen so
that heavy quarks reach thermal equilibrium in the long-time limit with
the temperature given by the surrounding medium, leading to the
Boltzmann-J\"uttner distribution,
\begin{equation}
\label{langevin.4}
f_{\text{eq}}(\vec{p}) \propto \exp(-\sqrt{\vec{p}^2+m^2}/T) \ .
\end{equation}
To avoid the evaluation of momentum derivatives of the diffusion
coefficients, the post-point Ito realization has been adopted for the
stochastic integral ($\xi=1$). This allows to set $\Gamma=A$ with the
drag coefficient given by Eq.~(\ref{drag-coefficient}). As for the
pre-point Ito, the longitudinal drag coefficient has been enforced to
obey the Einstein-dissipation-fluctuation relation, $B_L=T E A$, with
$E=\sqrt{m^2+\vec{p}^2}$, which indeed leads to the equilibrium limit,
Eq.~(\ref{langevin.4}), independently of the specific momentum
dependence of the drag coefficient $A$~\cite{Rapp:2009my}. For a flowing
background medium, first the momentum coordinates have been
Lorentz-boosted to the local heat-bath rest frame. After performing the
time step (\ref{langevin.1}) in this frame, the new momentum variables
have been transformed back to laboratory-frame coordinates. This
procedure leads to an equilibrium limit
\begin{equation}
\label{langevin.5}
f_{\text{eq}}(\vec{p}_{\text{lab}})  \propto \frac{p_{\text{lab}} \cdot
  u}{E_{\text{lab}}} e^{-\frac{p_{\rm lab}\cdot u}{T}} \ .
\end{equation}
In terms of a Cooper-Frye freeze-out description,
Eq.~(\ref{cooper-frye}), this corresponds to the choice of a
hyper-surface element $\dd \sigma^{\mu} = \dd^3 x \; u^{\mu}$, which is
similar to the modified Milekhin-freeze-out description discussed in
Refs.~\cite{Russkikh:2006aa,Ivanov:2008zi}. In order to render the
thermal-fireball description of the bulk consistent with the freeze-out
prescription implied by the equilibrium limit of the Langevin
realization, the elliptic fireball described in Sec.~\ref{sect:medium}
has been adjusted using Eq.~(\ref{langevin.5}) as the local-equilibrium
distribution of the light quarks. This leads to a light-quark bulk
elliptic flow of $v_2 \simeq 5.6\%$, as shown by the endpoint of the
Milekhin-like curve in Fig.~\ref{fig_incl_v2}. This value is compatible
with the experimental $v_2$ for light hadrons when applying the
coalescence model for hadronization of Ref.~\cite{Greco:2003mm} (the
same model has been subsequently used in
Refs.~\cite{vanHees:2005wb,vanHees:2007me} to convert the heavy-quark
spectra from the Langevin simulations into $D$- and $B$-meson
spectra). As pointed out in the text, if instead the standard CF
constant-lab time freeze-out prescription is used, for which $\dd \sigma
= \dd^3 x (1,0,0,0)$, the same fireball evolution leads to a higher bulk
elliptic flow of $v_2 \simeq 9.2 \%$ for light quarks of mass $m_q=300
\; \MeV$. This finding reiterates a main point of the present
investigation: Conclusions about the microscopic dynamics (i.e, the
transport coefficients of heavy quarks in the QGP) from experimental
heavy-quark observables, like $R_{AA}$ and $v_2$ of single electrons at
RHIC, depend on the description of the expansion of the QGP.  This
expansion has to be understood from the light-hadron data.

\end{document}